\documentclass[pra,preprint,showpacs]{revtex4}

\usepackage{graphicx}

\begin{document}

\title{Useful Alternative to the Multipole Expansion of $1/r$ Potentials}

\author{Howard S. Cohl}
\email{hcohl@datasync.com}
\altaffiliation[Present Address: ]
{Institute of Geophysics and Planetary Physics,
Lawrence Livermore National Laboratory,
Livermore, CA, 94550}
\altaffiliation
{Logicon, Inc., Naval Oceanographic Office Major Shared Resource
Center Programming Environment \& Training,
NASA John C. Stennis Space Center, MS, 39529.}
\author{A. R. P. Rau}
\author{Joel E. Tohline}
\author{Dana A. Browne}
\author{John E. Cazes}
\altaffiliation[Present address: ]
{Logicon, Inc., Naval Oceanographic Office Major Shared Resource
Center Programming Environment \& Training,
NASA John C. Stennis Space Center, MS, 39529.}
\author{Eric I. Barnes}
\affiliation{Department of Physics and Astronomy,
Louisiana State University, 
Baton Rouge, LA, 70803}

\date{\today}

\begin{abstract}

Few-body problems involving Coulomb or gravitational interactions
between pairs of particles, whether in classical or quantum physics,
are generally handled through a standard multipole expansion of the
two-body potentials.  We develop an alternative based on an old, but
hitherto forgotten, expression for the inverse distance between two
points that builds on azimuthal symmetry. This alternative should have
wide applicability throughout physics and astronomy, both for
computation and for the insights it provides through its emphasis on
different symmetries and structures than are familiar from the standard
treatment. We compare and contrast the two methods, develop new
addition theorems for Legendre functions of the second kind, and a
number of useful analytical expressions for these functions.
Two-electron ``direct'' and ``exchange'' integrals in many-electron
quantum systems are evaluated to illustrate the procedure which is more
compact than the standard one using Wigner coefficients and Slater
integrals.

\end{abstract}


\maketitle

\section{Introduction}

For pairwise Coulomb or gravitational potentials, one often expands the
inverse distance between two points $\mathbf{x}$ and $\mathbf{x^\prime}$ in 
the standard multipole form \cite{ref1}
\begin{equation}
\frac{1}{|\mathbf{x} - \mathbf{x^\prime}|}=\frac{1}{\sqrt{rr^\prime}}
\sum_{\ell=0}^\infty
\biggl(\frac{r_{<}}{r_{>}}\biggr)^{\ell+\frac{1}{2}}P_\ell(\cos\gamma),
\label{eqn1}
\end{equation}
where $r_{<}$ $(r_{>})$ is the smaller (larger) of the spherical
distances $r$ and $r^\prime$, and $P_\ell(\cos\gamma)$ is
the Legendre polynomial \cite{ref2} with argument 
\begin{equation}
   \cos\gamma \equiv \mathbf{{\hat{x}}}\cdot\mathbf{{\hat{x}}}^\prime =
   \cos\theta\cos\theta^\prime+
   \sin\theta\sin\theta^\prime\cos(\phi-\phi^\prime).
\label{eqn2}
\end{equation}
The set of six coordinates \{$\mathbf{x},\mathbf{x^\prime}$\} may be
viewed either as defining two points relative to the origin or as the
coordinates of a three-body system once the motion of the center of
mass has been separated. In the ``body frame'', three out of the six
coordinates are dynamical variables, the potential energy depending
only on them. Of various choices for these variables, one is the set of
three separation distances, another the triad ($r_{<}$, $r_{>}$, $\gamma$)
as in Eq.~(\ref{eqn1}).  With respect to a space-fixed ``laboratory
frame,'' three more angles constitute the full set of six coordinates,
the choice in Eq.~(\ref{eqn2}) of ($\theta$, $\theta^\prime$, $\phi -
\phi^\prime$) being suited to the spherical polar coordinates of the
individual vectors; thus, $\mathbf{x}$: ($r \sin \theta  \cos \phi$,
$r \sin \theta \sin \phi$, $r \cos \theta$).

In this paper, we present an alternative expansion to Eq.~(\ref{eqn1})
based on cylindrical (azimuthal) symmetry which should be of wide
interest in physics and astrophysics.  This expansion arose in a recent
investigation by two of us \cite{ref3} of gravitational potentials in
circular cylindrical coordinates
$\mathbf{x}$ =  ($R$, $\phi$, $z$):
\begin{equation}
\frac{1}{|\mathbf{x} - \mathbf{x^\prime}|}=
\frac{1}{\pi\sqrt{RR^\prime}}
\sum_{m=-\infty}^{\infty} 
\ Q_{m-\frac{1}{2}}(\chi)
\ e^{im(\phi-\phi^\prime)},
\label{eqn3}
\end{equation}
\noindent with $Q_{m-\frac{1}{2}}$ a Legendre 
function of the second kind of half-integer degree \cite{ref4}, and $\chi$ 
defined as
\begin{equation}
\chi\equiv\frac{R^2+R^{\prime^2}+(z-z^\prime)^2}{2RR^{\prime}}=
\frac{r^2+{r^\prime}^2-2rr^\prime\cos\theta\cos\theta^\prime}
{2rr^\prime\sin\theta\sin\theta^\prime}.
\label{eqn4}
\end{equation}

Although this expansion has been recorded in many places \cite{ref5,ref6},
its full significance has not been appreciated. We have traced its
earliest occurrence to the work of E. Heine in the mid-nineteenth
century \cite{ref5} and will, therefore, call it the ``Heine identity.''  At
its most general, it takes the form
\begin{equation}
\frac{1}{\sqrt{v - \cos \psi}} = \frac{\sqrt{2}}{\pi} 
\sum_{n=-\infty}^\infty Q_{n-\frac{1}{2}}(v)\  e^{in\psi},
\label{eqn5}
\end{equation}
reducing to Eq.~(\ref{eqn3}) when applied to the inverse distance
between two points. We have found it to be a more efficient approach
for problems with a cylindrical geometry and have used it for compact
numerical evaluation of gravitational potential fields of several
axisymmetric and nonaxisymmetric mass distributions \cite{ref3}. We now
set Eq.~(\ref{eqn3}) in a broader context, together with new associated
addition theorems and a novel application in quantum physics, hoping to
encourage wider use of this expansion throughout physics.

\section{The Alternative Expansions}

The expansion in Eq.~(\ref{eqn1}) disentangles the dynamics contained
in the radial variables from symmetries, particularly under rotations
and reflections, pertaining to the angle $\gamma$. Whereas the three
variables \{$r_{<},r_{>},\gamma $\} at this stage are joint coordinates
of $\mathbf{x}$ and $\mathbf{x^\prime}$, a further disentangling in
terms of the independent coordinates so as to handle permutational and
rotational symmetry aspects of the problem is often useful and achieved
through the addition theorem for spherical harmonics \cite{ref7}:
\begin{eqnarray}
P_\ell(\cos\gamma)&=&\frac{4\pi}{2\ell+1}\sum_{m=-\ell}^{\ell} Y_{\ell
m}(\theta,\phi) Y_{\ell m}^\ast(\theta^\prime,\phi^\prime)\nonumber\\
& = & \sum_{m=-\ell}^{\ell} \frac{\Gamma(\ell-m+1)}{\Gamma(\ell+m+1)}
P_\ell^m(\cos\theta) P_\ell^m(\cos\theta^\prime) e^{im(\phi-\phi^\prime)},
\label{eqn6}
\end{eqnarray}
where $Y_{\ell m}$ are the usual
spherical harmonics \cite{ref7}, $\Gamma(z)$ is the gamma function,
and $P_\ell^m(z)$ is the integer-order, integer-degree, associated
Legendre function of the first kind \cite{ref2}.  Using this to replace
$P_{\ell}(\cos\gamma)$ in Eq.~(\ref{eqn1}), we obtain the familiar
Green's function multipole expansion in terms of all six spherical polar
coordinate variables $\bf x$ and $\bf x^\prime$,
\begin{equation}
\frac{1}{|\mathbf{x} - \mathbf{x^\prime}|}=
\frac{1}{\sqrt{rr^\prime}}
\sum_{\ell=0}^\infty
\biggl(\frac{r_{<}}{r_{>}}\biggr)^{\ell+\frac{1}{2}}
\sum_{m=-\ell}^\ell
\frac{\Gamma(\ell-m+1)}{\Gamma(\ell+m+1)}
P_\ell^m(\cos\theta) P_\ell^m(\cos\theta^\prime)
e^{im(\phi-{\phi}^\prime)}.
\label{eqn7}
\end{equation}

Apart from the first factor with dimension inverse-distance formed from
the geometric mean of the two lengths $r$ and $r^\prime$, this expression
involves only four combinations: $(r_{<}/r_{>}, \theta, \theta^\prime, \phi -
\phi^\prime)$ of the six coordinates $\bf x$ and $\bf x^\prime$. In spite
of widespread familiarity with the multipole expansion, this reduction
from six to four essential variables has not been appreciated fully. We
were led to it by the parallel investigation below of the alternative
expansion, and note that it is a natural consequence of the separation
distance being independent of the orientation of that separation in the
laboratory frame, and thus independent of two angles serving to specify
that orientation.

This multipole expansion in terms of spherical harmonics is very broadly
utilized across the physical sciences.  For example, with $\ell$ and $m$
interpreted as the quantum numbers of orbital angular momentum and its
azimuthal projection, respectively, a whole technology of Racah-Wigner or
Clebsch-Gordan algebra is available \cite{ref8} for handling all angular
(that is, geometrical or symmetry) aspects of an $N$-body problem, the
dynamics being confined to radial matrix elements of the coefficients
$(r_{<}/r_{>})^{\ell+\frac{1}{2}}$ in Eq.~(\ref{eqn1}).  Although many other
systems of coordinates have been studied for problems with an underlying
symmetry that is different from the spherical, Eq.~(\ref{eqn1}), in
combination with the addition theorem for spherical harmonics,  has
gained such prominence as to have become the Green's function expansion
of choice even for non-spherically symmetric situations.

\begin{figure}
\includegraphics[width=3in]{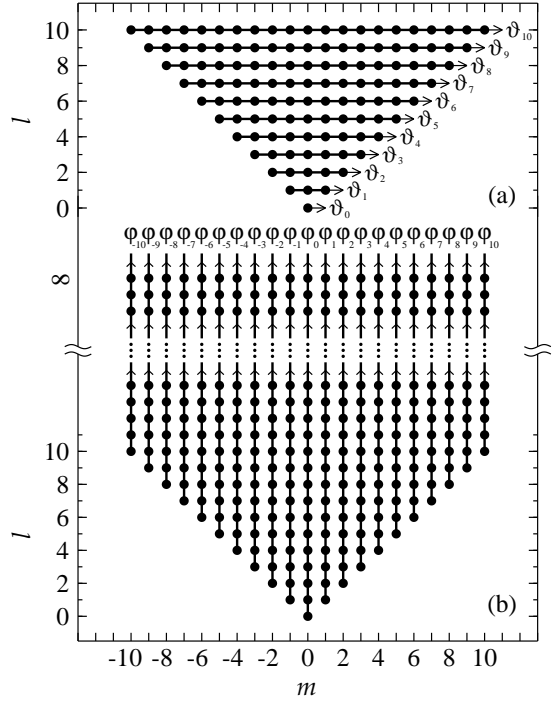}
\caption{Alternative double summations in ($\ell$,$m$) space (a) first 
over $m$ at fixed $\ell$ to form partial sums $\vartheta_\ell$
as in Eq.~(\ref{eqn7})
(b) first over $\ell$ at fixed $m$ to form partial sums $\varphi_m$
as in Eq.~(\ref{eqn8})}
\label{fig1}
\end{figure}

But, consider now the expansion in Eq.~(\ref{eqn3}), which may be
viewed either, in analogy with Eq.~(\ref{eqn1}), as an expansion in
Legendre functions, now of the second kind in the joint variable $\chi$
of the whole system, with coefficients $(RR^\prime)^{-1/2}e^{im(\phi
-\phi^\prime)}$, or as a Fourier expansion in the variable $(\phi
-\phi^\prime)$ with the $Q$'s as coefficients.  In this latter view,
a further step allows us to develop a new addition theorem for these
Legendre functions.  Interchanging the $\ell$ and $m$ summations
(Fig.~\ref{fig1}) in Eq.~(\ref{eqn7}), we obtain
\begin{equation}
\frac{1}{|\mathbf{x} - \mathbf{x^\prime}|}=
\frac{1}{\sqrt{rr^\prime}}
\sum_{m=-\infty}^{\infty} 
e^{im(\phi-\phi^\prime)}
\sum_{\ell=|m|}^{\infty} 
\biggl(\frac{r_{<}}{r_{>}}\biggr)^{\ell+\frac{1}{2}}
\frac{\Gamma(\ell-m+1)}{\Gamma(\ell+m+1)}
P_{\ell}^{m}(\cos\theta)
P_{\ell}^{m}(\cos\theta^\prime).
\label{eqn8}
\end{equation}
Comparing with Eq.~(\ref{eqn3}), we deduce: 
\begin{equation}
Q_{m-\frac{1}{2}}(\chi)
=\pi\sqrt{\sin\theta \sin\theta^\prime}
\sum_{\ell=|m|}^{\infty} 
\biggl( \frac{r_{<}}{r_{>}} \biggr)^{\ell+\frac{1}{2}}
\frac{\Gamma(\ell-m+1)}{\Gamma(\ell+m+1)}
P_{\ell}^{m}(\cos\theta)
P_{\ell}^{m}(\cos\theta^\prime).
\label{eqn9}
\end{equation}
This is a new addition theorem for the Legendre function of the second kind.
Note that $Q_{m-\frac{1}{2}}=Q_{-m-\frac{1}{2}}$ as per Eq.~(8.2.2)
in \cite{ref2}.

Similarities and contrasts between the pairs of equations,
Eqs.~(\ref{eqn1}) and (\ref{eqn7}) and Eqs.~(\ref{eqn3}) and
(\ref{eqn8}), are worth emphasizing. Of the four variables $(r_{<}/r_{>},
\theta, \theta^\prime, \phi -\phi^\prime)$, the first pair of equations,
Eqs.~(\ref{eqn1}) and (\ref{eqn7}), expresses the inverse distance as
a series in powers of the first variable with coefficients Legendre
polynomials of the first kind in $\gamma$, which is a composite of the
other three variables and decomposable in terms of them through the
addition theorem as in Eq.~(\ref{eqn6}). The second pair of equations,
Eqs.~(\ref{eqn3}) and (\ref{eqn8}), on the other hand, expands in
Eq.~(\ref{eqn3}) the inverse distance in terms of the variable $\phi
- \phi^\prime$, with expansion coefficients Legendre functions of the
second kind in $\chi$, which is a composite of the other three variables
$(r_{<}/r_{>}, \theta, \theta^ \prime)$ and decomposable in terms of them
through the addition theorem in Eq.~(\ref{eqn9}). For this comparison,
it is useful to recast Eq.~(\ref{eqn1}) in the more suggestive form,
\begin{equation}
\frac{1}{|\mathbf{x}-\mathbf{x^\prime}|}=
\frac{1}{\sqrt{rr^\prime}}\sum_{\ell=0}^\infty
P_{\ell}(\cos\gamma)\  e^{-(\ell+\frac{1}{2})(\ln r_{>} -\ln r_{<})}.
\label{eqn10}
\end{equation}
Whereas this expansion has half-integers in the exponents and integer
degree Legendre polynomials of the first kind as coefficients,
Eq.~(\ref{eqn3}) has integer $m$'s in the exponents and half-integer
degree Legendre functions of the second kind as coefficients.

Yet another alternative to Eq.~(\ref{eqn3}) follows upon casting the
square root in the expression for the distance in terms of $r, r^\prime,$
and $\gamma$ in the form Eq.~(\ref{eqn5}) through the definition
\begin{equation}
v \equiv \frac{1}{2}\biggl( \frac{r_{<}}{r_{>}} + \frac{r_{>}}{r_{<}} \biggr)
= \frac{r^2+{r^\prime}^2}{2rr^\prime}.
\label{eqn11}
\end{equation}
This gives the expression 
\begin{equation}
\frac{1}{|\mathbf{x} - \mathbf{x^\prime}|}=
\frac{1}{\pi\sqrt{rr^\prime}}\sum_{n=-\infty}^\infty
Q_{n-\frac{1}{2}}(v)\  e^{in\gamma},
\label{eqn12}
\end{equation}
now a Fourier expansion in $\gamma$ instead of the angle
$(\phi-\phi^\prime)$ of Eq.~(\ref{eqn3}), with $Q_{n-\frac{1}{2}}(v)$ as the
coefficients. In terms of hyperspherical coordinates, widely used in
atomic and nuclear study of three (or more) bodies \cite{ref9}, the
variable $v$ is $\csc 2\alpha$, $\alpha$ being a ``hyperangle''. In an
Appendix, we present a number of alternative expressions for the functions
$Q_{m-\frac{1}{2}}$, some of them new, which we have found useful while
working with the expansions in Eqs. (\ref{eqn3}) and (\ref{eqn12}).

\section{Two-electron integrals}

One familiar application of expressions for the inverse distance is
to the Coulomb interaction between two charges. We contrast usage
of the alternative expansions in Eqs.~(\ref{eqn1}) and (\ref{eqn3})
for calculating the electrostatic interaction as it appears in atomic,
molecular and condensed matter physics. Thus, consider the so-called
``direct'' part of this interaction between two electrons in the $3d^2$
configuration,
\begin{equation}
V_{ee}^D=\int d\mathbf{x}\int d\mathbf{x'}
\psi_{3d}^*(\mathbf{x}) \psi_{3d}^*(\mathbf{x '})
\frac{1}{|\mathbf{x}-\mathbf{x'}|}
\psi_{3d}(\mathbf{x})\psi_{3d}(\mathbf{x '}).
\label{eqn13}
\end{equation}
The standard treatment \cite{ref10} uses Eqs.~(\ref{eqn1}) and
(\ref{eqn8}), carries out all the angular integrals through Racah-Wigner
algebra, leaving behind radial ``Slater integrals'' $F^k (dd), k=0,2,4,$
and yielding (for illustrative purposes, all $m$ values have been set
equal to zero)
\begin{equation}
V_{ee}^D=F^0 (dd) + (4/49)F^2 (dd) + (36/441)F^4 (dd),
\label{eqn14}
\end{equation}
where the coefficients are evaluated in terms of Wigner $3j$-symbols 
or are available in tables \cite{ref10}. The Slater integrals,
\begin{equation}
F^k (dd)=\int \int r^2 dr r'^2 dr' (r_{<}^k/r_{>}^{k+1}) R_{3d}
^2 (r) R_{3d}^2 (r'),
\label{eqn15}
\end{equation}
remain for numerical evaluation. In this example, upon evaluation
with hydrogenic radial functions, we obtain $V=0.092172$ in atomic units.

The alternative calculation through Eq.~(\ref{eqn3}) involves only the $m=0$
term and thereby the integral
\begin{equation}
V_{ee}^D=\frac {1}{\pi}\int\int\int\int R^{1/2}dRR'^{1/2}dR'dzdz'
Q_{-\frac {1}{2}} (\chi)|\psi_{3d}(\mathbf{x})|^2 |\psi_{3d}(\mathbf{x'})|^2.
\label{eqn16}
\end{equation}
The integrand is a function of $z$ and $R$ variables alone and our 
numerical evaluation of this integral reproduces the value cited above.

As a second example, we computed an exchange integral for the $3d4f$
configuration again setting, for simplicity, all $m$ equal to zero:
\begin{equation}
V_{ee}^E=\int\int d\mathbf{x}d\mathbf{x'}\psi_{3d}^*(\mathbf{x})
\psi_{4f}^*(\mathbf{x'})\frac{1}{|\mathbf{x}-\mathbf{x'}|}
\psi_{3d}(\mathbf{x'})\psi_{4f}(\mathbf{x}).
\label{eqn17}
\end{equation}
The standard method through exchange Slater integrals $G$ and Wigner
coefficients gives \cite{ref10}
\begin{equation}
V_{ee}^E=(9/35)G^1 (df)+(16/315)G^3 (df)+(500/7623)G^5 (df)
\label{eqn18}
\end{equation}
and, again through hydrogenic radial functions, gives the value
$V_{ee}^E=0.0082862$. We reproduce the same result upon directly
computing Eq.~(\ref{eqn17}) with Eq.~(\ref{eqn3}), again involving a
single four-dimensional integral as in Eq.~(\ref{eqn16}) with
$Q_{-\frac{1}{2}}$.

As the orbital angular momenta involved of the two electrons increase,
the number of terms in expressions such as Eqs.~(\ref{eqn14}) and
(\ref{eqn18}) also grows, necessitating the computing of more Wigner
coefficients and Slater integrals. By contrast, only a single term of
the expansion in Eq.~(\ref{eqn3}) and a single integral is necessary
in our suggested alternative, the $\phi$ integrations setting $m=0$
for direct terms and $m$ equal to the difference in the $m$ values
of the two orbitals for exchange terms.  This same selection rule
imposed by the $\phi$ integrations means that even in a calculation
with several configurations and the imposition of antisymmetrization,
such as in a multi-configuration Hartree-Fock scheme, matrix elements
of $|\mathbf{x} -\mathbf{x'}|^{-1}$ between each term in the bra and in
the ket gets a contribution from only one $m$ value in the expansion in
Eq.~(\ref{eqn3}). This is a significant economy.

Although the evaluation of the four-dimensional integrals in
Eq.~(\ref{eqn16}) is computationally more demanding than the
two-dimensional integrals of Eq.~(\ref{eqn15}), the preparatory work
of the Wigner--Racah algebra is avoided.  A more efficient alternative
to directly evaluating the four-dimensional integrals in expressions
such as Eq.~(\ref{eqn16}) is to use the fact that they can be viewed as
electrostatic interaction energies between charge densities given by the
product of wave functions.  The charge density $\rho(\mathbf{x'})$ is then
used to find the potential at $\mathbf{x}$ and the resulting potential
integrated with the appropriate charge density $\rho(\mathbf{x})$ to
compute the integral \cite{ref11}.  This same approach applies to the
decomposition according to Eq.~(\ref{eqn3}) for each $m$-th component of
the potential. A two-dimensional Poisson equation in the primed variables
in Eq.~(\ref{eqn16}) is first solved and then Eqs.~(\ref{eqn16}) and
(\ref{eqn17}) evaluated as two-dimensional integrals over the unprimed
variables \cite{ref12}. As already noted, direct integrals will involve
only the $m=0$ component; exchange ones involve a single $m$ value equal
to the difference in the azimuthal quantum numbers of the two orbitals.

\section{Summary}

The inverse distance between two points $\mathbf{x^\prime}$ and
$\mathbf{x}$ is intimately involved in Coulomb and gravitational
problems. Its expansion in terms of Legendre polynomials $P_\ell$
of the angle between the vector pair or a further double-summation
expansion involving the individual polar angles of the vectors are well
known and widely used in physics and astronomy. We have discussed an
alternative in terms of cylindrical coordinates, a single summation in
terms of Legendre functions $Q_{m-\frac {1}{2}}$ of the second kind in
a pair variable $\chi$ or double summations involving the individual
coordinates. These expansions are better suited to problems that are
decomposable in azimuthal symmetry as shown by applications in \cite{ref3}
and by an illustration here for very common electron-electron calculations
throughout many-electron physics.  Further variants are possible for
other coordinates such as ring or toroidal, parabolic, bispherical,
cyclidic and spheroidal \cite{ref13}, and we hope to return to them in
future publications.  Connections to the theory of Lie groups will also
be of interest \cite{ref14}.

\begin{acknowledgments}

We thank Prof. Ken Schafer for his remark that better ways of
evaluating two-electron exchange integrals would be useful.  ARPR
thanks the Alexander von Humboldt Stiftung and Profs. J. Hinze and F.
H. M. Faisal of the University of Bielefeld for their hospitality
during the course of this work. This work has been supported, in part,
by NSF grant AST-9987344 and NASA grant NAG5-8497. Work performed at
LLNL is supported by the DOE under Contract W7405-ENG-48.

\end{acknowledgments}

\appendix*

\section{Alternative expressions}

We present in this appendix a number of alternative expressions for the
functions $Q_{m-\frac{1}{2}}$ which are useful in calculations using
expansions such as Eqs.~(\ref{eqn3}) and (\ref{eqn13}). Setting $\theta
= \theta^\prime = \pi/2$ in Eq.~(\ref{eqn9}) and using Eq.~(8.756.1)
of \cite{ref15} gives
\begin{equation}
Q_{m-\frac{1}{2}}(v) = \pi \sum_{\ell=|m|}^{\infty}
\biggl(\frac{r_{<}}{r_{>}}\biggr)^{\ell+\frac{1}{2}} 
\frac{\Gamma(\ell-m+1)}{\Gamma(\ell+m+1)}
\frac{\pi 2^{2m}}{[\Gamma(1+\frac{\ell-m}{2})\Gamma(\frac{1-\ell-m}{2})]^2},
\label{eqn19}
\end{equation}
which can be rewritten as
\begin{equation}
Q_{m-\frac{1}{2}}(v) = \pi e^{-(m+\frac{1}{2})\eta}
\sum_{\ell =0}^\infty 2^{1-2m-4\ell} 
\biggl( \begin{array}{c} 2\ell \\ \ell \end{array} \biggr)
\biggl( \begin{array}{c} 2\ell +2m-1 \\ \ell +m \end{array} \biggr) e^{-2\ell
\eta},
\label{eqn20}
\end{equation}
where we have defined $r_{<}/r_{>}\equiv e^{-\eta}$, $v=\cosh\eta$.  Although
the $\ell$-th term of these series is in different form from what one
obtains through the more familiar formula for $Q$ as a hypergeometric
function \cite{ref16}, namely,
\begin{equation}
Q_{m-\frac{1}{2}}(v=\cosh\eta)= \frac{\sqrt{\pi}\Gamma(m+\frac{1}{2})}
{\Gamma(m+1)} e^{-(m+\frac{1}{2})\eta}
\ _2F_1(\frac{1}{2},m+\frac{1}{2};m+1;e^{-2\eta}),
\label{eqn21}
\end{equation}
their equivalence follows from straightforward algebra. Also, another
standard expansion for $Q$ in powers of $v$ as in Eq.~(8.1.3) of
\cite{ref2},
\begin{equation}
Q_{\nu-\frac{1}{2}}(v)=
\frac{\sqrt{\pi}\Gamma(\nu+\frac{1}{2})}{\Gamma(\nu +1)}(2v)^{-\nu-\frac{1}{2}} 
\ 
_2F_1(\frac{\nu}{2}+\frac{3}{4},\frac{\nu}{2}+\frac{1}{4};\nu+1;\frac{1}{v^2}),
\label{eqn22}
\end{equation}
is equivalent. However, the results directly in powers of $r_{<}/r_{>}$
in Eqs.~(\ref{eqn19}), (\ref{eqn20}), and (\ref{eqn21}) are more
convenient in many applications.  Among specific features worth noting
in these alternative expansions are that only even powers of $(r_{<}/r_{>})$
occur in the sum in Eq.~(\ref{eqn20}) and that for any $m$, the sum
in Eq.~(\ref{eqn19}) runs over all $\ell$ values compatible with it,
$\ell \geq |m|$, as per their interpretation as angular momentum quantum
numbers.

In the multipole expansion in Eq.~(\ref{eqn1}), $\gamma$ is an angle
formed out of the set $(\theta, \theta^\prime, \phi -\phi^\prime)$ and,
therefore, $\cos\gamma$ in the functions $P_\ell$ has range of variation
from --1 to 1. On the other hand, in the expansions in Eqs.~(\ref{eqn3})
and (\ref{eqn5}), the arguments $v$ and $\chi$ of the Legendre functions
of the second kind range from 1 to infinity and, therefore, can be
written in terms of hyperbolic functions as $\cosh\eta$ and $\cosh\xi$,
respectively.  From Eq.~(\ref{eqn4}) we have the link between them,
\begin{equation}
\cosh\eta = \cos \theta\cos \theta^\prime
+\sin \theta\sin \theta^\prime \cosh \xi.
\label{eqn23}
\end{equation}
This disentanglement of $v$ (or $\eta$) in terms of a triad is the
counterpart of Eq.~(\ref{eqn2}) and may be used with addition theorems
given in the literature such as \cite{ref5,ref6}
\begin{equation}
Q_{m-\frac{1}{2}}(\cosh\eta) = \sum_{n=-\infty}^\infty  (-1)^n
\frac{\Gamma(m-n-\frac{1}{2})}{\Gamma(n+m-\frac{1}{2})}
Q_{m-\frac{1}{2}}^n(\cos\theta) 
P_{m-\frac{1}{2}}^n(\cos\theta^\prime)
\ e^{n\xi}.
\label{eqn24}
\end{equation}
An alternative to writing $\chi$ as $\cosh\xi$ is to set $\chi=\coth\zeta$
in Eq.~(\ref{eqn23}). In that case, it follows that
\begin{equation}
\exp\zeta\equiv
\sqrt{
\frac
{\cosh\eta-\cos(\theta+\theta^\prime)}
{\cosh\eta-\cos(\theta-\theta^\prime)}
}.
\label{eqn25}
\end{equation}

Other variants of the expansions in Eqs.~(\ref{eqn9}) and (\ref{eqn24})
follow from identities satisfied by the Legendre functions $P$ and
$Q$. In particular, there is an interesting pair of relations involving
index interchange given in Eqs.~(8.2.7) and (8.2.8) of \cite{ref2}:
\begin{eqnarray}
Q_{n-\frac{1}{2}}^m(\cosh\eta)=(-1)^m
\sqrt{\frac{\pi}{2\sinh\eta}}
\Gamma(m-n+\frac{1}{2})
\ P_{m-\frac{1}{2}}^n(\coth\eta),\\
Q_{m-\frac{1}{2}}^n(\coth\eta)
=(-1)^m
\sqrt{\frac{\pi\sinh\eta}{2}}
\frac{\pi}{\Gamma(m-n+\frac{1}{2})}
P_{n-\frac{1}{2}}^m(\cosh\eta).
\end{eqnarray}

\end{document}